# Thermal Interaction-Free Ghost Imaging

Shun Li,[1] Jing-Yang Xiao Feng,[1] Xiu-Qing Yang,[2,*] Xiaodong Zeng,[1,3] Xi-Hua Yang,[1,3] M. Al-Amri,[4,5,*] and Zheng-Hong Li[1,3,*]

*1 Department of Physics and Institute for Quantum Science and Technology, Shanghai University, Shanghai 200444, China.*

*2 College of Science, Inner Mongolia University of Technology, Hohhot 010051, China*

*3 Shanghai Key Laboratory of High Temperature Superconductors, Shanghai University, Shanghai 200444, China*

*4 NCQOQI, KACST, P. O. Box 6086, Riyadh11442, Saudi Arabia*

*5 Institute of Quantum Technologies and Advanced Computing, KACST, Riyadh 11442, Saudi Arabia*

## ABSTRACT

We propose an interaction-free ghost imaging scheme based on a thermal light source. By utilizing the quantum Zeno-like effect, our approach significantly reduces the light dose absorbed by the sample, thereby effectively preventing sample damage induced by light-matter interactions. Combined with the elimination of entangled photon sources and single-photon detectors, our approach enables significantly more photons to be utilized for image reconstruction, thereby markedly enhancing image quality compared to conventional ghost imaging. We further demonstrate active suppression of background noise via controllable photon loss. Our work offers a practical and cost-effective route to non-destructive, high-quality imaging for light-sensitive samples in fields such as life sciences.

## I. INTRODUCTION

Increasing the total photon number in the illumination field is one of the most straightforward methods to enhance image quality. However, this approach is limited by practical constraints such as light-induced sample damage. These limitations have motivated research into weak-field imaging, with quantum ghost imaging serving as a prominent example [1,2]. It utilizes non-classical correlations between entangled photon pairs and employs coincidence measurements to suppress random noise, enabling effective extraction of object information from extremely weak optical fields [3-5]. Although the image quality in quantum ghost imaging still benefits from photon accumulation, the low generation rate of entangled photon pairs makes it impractical to significantly increase the total photon number. Moreover, the low-count-rate of single-photon detectors further limits imaging speed, resulting in an inverse relationship between acquisition time and photon count [5-7].

In addition to weak-field imaging, quantum mechanics also permits a more fundamental approach to avoid damage caused by light-matter interactions: interaction-free measurement [8-14]. Combined with the quantum Zeno effect, interaction-free measurement theoretically allows information acquisition without any physical “contact” [15-20]. This capability not only stimulates theoretical exploration of nonlocal quantum phenomena [21-24] but also holds promise for significant applications in fields such as quantum communication [25-29], quantum information processing [30-34], and quantum sensing [35-40]. Particularly for light-sensitive samples, this method offers a powerful tool for the dynamic observation of biological specimens like living cells and proteins [12,15,39]. Recently, researchers have begun exploring the combination of interaction-free measurement with quantum ghost imaging [41-43]. However, such hybrid schemes still require entangled sources and single-photon detectors, and thus fail to overcome the efficiency bottleneck inherent in quantum ghost imaging.

It is noteworthy that ghost imaging does not inherently require quantum light [44-47]. Initial work in 2002 demonstrated that ghost imaging could be performed without entanglement [48]. Building on this, a practical scheme utilizing thermal light was subsequently proposed, which operated without the need for single-photon detectors [49]. Compared to quantum ghost imaging, thermal ghost imaging consistently suffers from substantial background noise, meaning that, under the same total photon number in the illumination field, it exhibits inferior image quality [6]. However, thermal ghost imaging has its own advantages. We note that the image quality of ghost imaging is proportional to both the number of measurements (denoted as $K$) and the average photon number per measurement. Studies also indicate that, in thermal ghost imaging, the photon count per measurement can be very high, whereas quantum ghost imaging typically uses only one photon per measurement. This allows thermal ghost imaging to achieve better image quality in a single measurement while also offering a relatively faster imaging speed [6]. In general, the two approaches are suited to different scenarios. The quantum method is more appropriate for situations with limited total photon numbers or extremely weak illumination (where photons in the illumination field are a scarce resource). The thermal method holds an advantage when strong illumination is permissible (where photons are abundant), and its main challenges lie in the increased risk of sample damage under high light doses, as well as inherent background noise [50].

In this work, we combine, for the first time, interaction-free measurement with thermal ghost imaging, dramatically reducing light-induced damage. This allows our approach to safely increase the number of measurements to levels unattainable by traditional thermal ghost imaging, thereby significantly improving the image quality. The study further reveals that optical loss can serve as an effective parameter for regulating background noise, offering a new dimension for actively optimizing imaging performance. Moreover, as a result of eliminating single-photon detectors and entangled sources, our scheme enables faster imaging, higher system stability, and lower cost, in contrast to quantum ghost imaging schemes that employ interaction-free measurement. It is also worth mentioning that our theoretical framework is applicable to computational ghost imaging [51-57], which evolved from thermal ghost imaging and operates on the same underlying principles.

## II. MODEL AND THEORY

### A. System configuration and imaging signal definition

Fig. 1(a) presents the thermal light interaction-free ghost imaging scheme. Our system introduces a chain interferometer structure into the signal path of the conventional approach [50], enabling multiple "interactions" between the illumination field and the sample. Thanks to this configuration, the light that would be directly absorbed by the sample in traditional schemes is now collected and directed to the non-spatially resolving bucket detector $D_0$, while the light transmitted through the transparent regions of the sample is guided to bucket detector $D_1$. The reference path is separated from the signal path by a 50:50 beam splitter $BS_H^{(1)}$, while a second $BS_H^{(2)}$ evenly divides the reference light into two beams, which are then respectively sent to two $CCD$s. These undergo coincidence measurements with the two bucket detectors, and the final measurement signal is obtained through differential output.

The chain structure consists of mirrors ($Mr$) and $M$ beam splitters $BS_M^{(m)}$ with identical reflectivity ($\cos^2(\pi/2M)$), where the superscript $m$ denotes the $m$-th beam splitter. When $M = 1$, the system reduces to the conventional thermal light ghost imaging configuration [50]. Fig. 1(b) shows the equivalent Michelson interferometer version [34] corresponding to Fig. 1(a). By using switchable mirrors ($SM$), the light pulses are confined within the interferometer cavity [58], forced to interact with the sample $M$ times in cycles before being released to the detectors. While Fig. 1(b) depicts a more practical application scenario [34], the configuration in Fig. 1(a) is better suited for theoretical analysis and experimental verification. Therefore, this paper will use Fig. 1(a) as the basis for discussion.

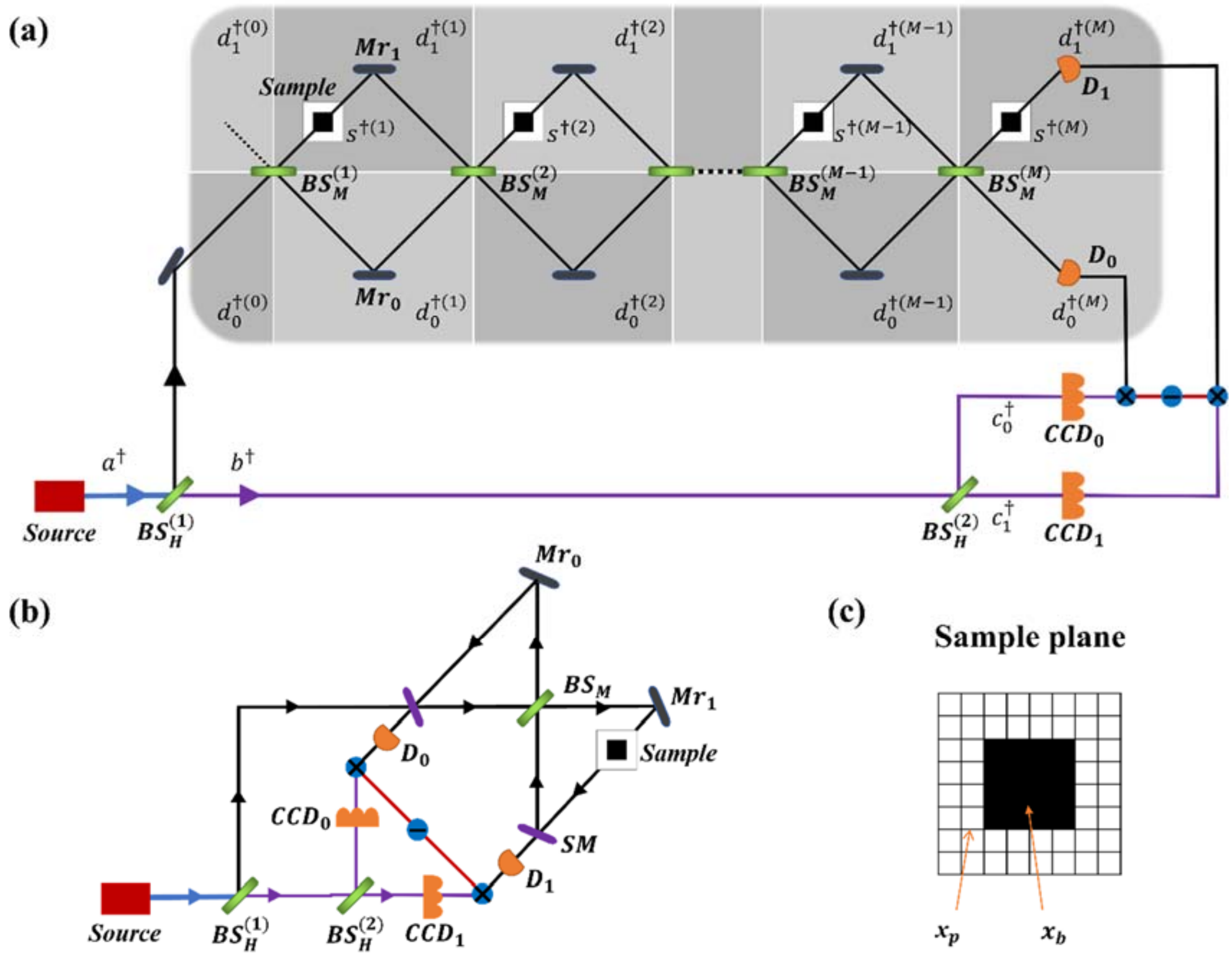

Fig. 1 Schematic of the thermal light interaction-free ghost imaging: (a) Implementation based on a chain interferometer structure; (b) Equivalent Michelson interferometer configuration; (c) Discretized model of the sample plane.

For simplicity, we focus on an ideal binary sample. As shown in Fig. 1(c), the sample surface is divided into $J$ units, each uniquely identified by $x_j$ $(j = 1,2, \dots, J)$, forming the set $\boldsymbol{J} = \{x_j\}$. Each unit is either transparent (denoted as $x_p$, belonging to set $\boldsymbol{J_p}$ with total number $J_p$) or opaque (denoted as $x_b$, belonging to set $\boldsymbol{J_b}$ with total number $J_b$). With the definition $\alpha = J_b/J_p$, the average transmittance of the sample is $1/(1+\alpha)$. Furthermore, we assume that each sample unit matches the size of an $CCD$ pixel and the average speckle size of the light source [50]. Under this configuration, the corresponding positions in the sample plane, detection plane, and source plane maintain a one-to-one spatial correspondence, all describable by the same coordinate $x$.

The optical field operators and measurement processes in the system are defined as follows. Let $a^\dagger$ denote the creation operator of the illumination field at the source plane. We assume that the light intensities in different speckles are statistically independent. The average number of photons within a single speckle per measurement is $2u$, i.e., $\langle I_a(x)\rangle = 2u$, where $I_a(x) = a^\dagger(x)a(x)$. The reference beam is described by $b^\dagger$. After passing through $BS_H^{(2)}$, the creation operators of the optical fields entering the two $CCD$s are denoted $c_i^\dagger$ with $i = 0,1$. Correspondingly, the average number of photons detected at pixel $x$ of $CCD_i$ per measurement is $I_{ci}(x) = c_i^\dagger(x)c_i(x)$. In the chain structure of the signal path, $d_1^{\dagger(m)}$ and $d_0^{\dagger(m)}$ represent the creation operators of the optical fields in the upper and lower paths, respectively, between the $m$-th beam splitter $BS_M^{(m)}$ and the next one. The average number of photons received by bucket detector $D_i$ ( $i = 0,1$ ) per measurement is $I_i = \sum_{x\in J} d_i^{\dagger(M)}(x)d_i^{(M)}(x)$. Additionally, $s^{\dagger(m)}$ represents the operator associated with light absorption by the sample during the $m$-th interaction.

After $K$ independent coincidence measurements between $D_i$ and $CCD_i$, their statistical average is denoted as $E[I_iI_{ci}(x)]$. Here, the expectation operator $E[\cdot]$ is defined as the arithmetic mean of any observable $X$, i.e., $E[X] = \sum_{k=1}^{K} X^{(k)}/K$, where $X^{(k)}$ is the value of $X$ in the $k$-th measurement. It should be noted that background noise is a non-negligible factor in thermal ghost imaging [48,49]. Here we adopt the most commonly used background subtraction method, directly subtracting the background noise term $E[I_i]E[I_{ci}(x)]$ [59-62]. The corrected coincidence measurement result can be expressed as $g_i(x) = E[I_iI_{ci}(x)] - E[I_i]E[I_{ci}(x)]$. Finally, the output signal $G$ of our scheme is defined as the difference between the corrected results from the two detection channels:

$$G(x) = g_1(x) - g_0(x). \tag{1}$$

### B. Calculation of the expected imaging signal $\langle G(x)\rangle$

First, we clarify the relations between the detection-end operators $d_i^{\dagger(M)}$ and $c_i^\dagger$, and the source operator $a^\dagger$. For the $CCD$, it is straightforward to obtain $c_i^\dagger(x) = a^\dagger(x)/2$ due to the two 50:50 $BS_H$s. For the bucket detectors, we define coefficients $\chi_{pi}$ and $\chi_{bi}$ such that $d_i^{\dagger(M)}(x_p) = \chi_{pi}a^\dagger(x_p)/\sqrt{2}$ for transparent units, while for opaque units, $d_i^{\dagger(M)}(x_b) =$ $\chi_{bi} a^{\dagger}(x_b)/\sqrt{2}$. The factor $1/\sqrt{2}$ originates from $BS_H^{(1)}$, while $\chi_{pi}$ and $\chi_{bi}$ depend on all optical components in the chain. Regarding $BS_M^{(m)}$, the input and output field operators satisfy the following relations: $d_0^{\dagger(m)} = d_0^{\dagger(m-1)} \cos\theta - d_1^{\dagger(m-1)} \sin\theta$ and $d_1^{\dagger(m)} = d_0^{\dagger(m-1)} \sin\theta + d_1^{\dagger(m-1)} \cos\theta$ where $\theta = \pi/(2M)$ [24]. As for the sample, transparent units leave the optical field unchanged, resulting in $d_1^{\dagger(m)} \to d_1^{\dagger(m)}$, while opaque units absorb and annihilate photons, leading to $d_1^{\dagger(m)} \to 0$. Under the condition that all other components in the optical path, such as the mirrors ($Mr$), are ideal, we obtain $\chi_{p0} = 0$, $\chi_{p1} = 1$, $\chi_{b1} = 0$, and $\chi_{b0} = \cos^M\theta$ [24] (The details can also be found in Appendix B). As $M$ approaches infinity, $\chi_{b0}$ approaches 1. In our calculations, we have assumed that the incident field is sufficiently intense (i.e., $u$ is very large) such that shot noise can be neglected [50], and similarly, the effects of vacuum input fields are also negligible (see Appendix A). This high-intensity illumination approximation is applied in all subsequent calculations.

When the mirrors are non-ideal, we assume that the reflectivity of each $Mr_0$ in the lower path is $\eta_0^2$, and the reflectivity of the $Mr_1$ in the upper path is $\eta_1^2$, meaning the loss probabilities per reflection are $\gamma_0 = 1 - \eta_0^2$ and $\gamma_1 = 1 - \eta_1^2$, respectively. In this case, $\chi_{b1} = 0$ and $\chi_{b0} = \eta_0^M \cos^M\theta$ (see Appendix B). However, $\chi_{p0}$ and $\chi_{p1}$ depend on $\eta_0$ and $\eta_1$ in a complex manner, which is better analyzed using numerical simulations. It should be noted that other optical losses in the chain structure can be uniformly modeled as the optical loss caused by $Mr_0$ and $Mr_1$.

After establishing the operator relations between the detection and source ends, we apply the Gaussian moment theorem [48,63] to calculate $\langle G(x) \rangle$. Under high-intensity illumination approximation, the expression of the theorem simplifies to: $\langle I_a^n(x) \rangle = n!\,(2u)^n$ and $\langle I_a(x) I_a(x') \rangle = \langle I_a(x) \rangle \langle I_a(x') \rangle = (2u)^2$, for $x \neq x'$ [50]. This yields the following results (see Appendix C):

$$\left\langle G\left(x_p\right)\right\rangle = \frac{K-1}{2K} C_p u^2, \quad \left\langle G\left(x_b\right)\right\rangle = -\frac{K-1}{2K} C_b u^2, \tag{2}$$

where $C_p = \left|\chi_{p1}\right|^2 - \left|\chi_{p0}\right|^2$ and $C_b = |\chi_{b0}|^2 - |\chi_{b1}|^2$. When all mirrors are perfect, with $K \gg 1$ and $M \gg 1$, we have $\left\langle G\left(x_p\right)\right\rangle = -\langle G(x_b) \rangle = u^2/2$. This confirms that our scheme effectively extracts sample information.

### C. Contrast-to-noise ratio (CNR)

CNR is a commonly used metric for evaluating ghost imaging performance [41-43,50], defined as: $\mathrm{CNR} \equiv \left(\langle G(x_p)\rangle - \langle G(x_b)\rangle\right)/\sqrt{\frac{1}{2}\left[\delta^2 G(x_p) + \delta^2 G(x_b)\right]}$, where $\delta^2 G \equiv \langle G^2\rangle - \langle G\rangle^2$. The numerator of this expression represents the unnormalized visibility, thus CNR can be regarded as a comprehensive measure combining signal-to-noise ratio and visibility. In Appendix C, we provide the derivation process, ultimately obtaining:

$$\mathrm{CNR} = \frac{\left(C_p + C_b\right)\sqrt{K-1}}{\sqrt{C_p^2\left(J_p + \frac{7}{2} - \frac{3}{K}\right) + C_b^2\left(J_b + \frac{7}{2} - \frac{3}{K}\right)}}. \tag{3}$$

It is worth noting that under the condition of $M = 1$ (with optical losses neglected, i.e., $\eta_0 = \eta_1 = 1$), we have $C_{\mathrm{p}} = 1$ and $C_b = 0$. In this case, Eq. (3) reduces to the known result in the traditional scheme [50].

Without loss of generality, consider the scenario where $K$, $J_{\mathrm{p}}$, and $J_b$ are all large. The CNR of traditional thermal ghost imaging is approximately $\mathrm{CNR}' \approx \sqrt{K'/J_p}$, while in our scheme, $\mathrm{CNR} \approx \left(2/\sqrt{1+\alpha}\right)\sqrt{K/J_p}$ when $M$ is sufficiently large. We note that the total number of photons in the illumination field is determined by both $u$ and $K$. Although $u$ is the key factor enabling thermal ghost imaging to achieve a higher imaging speed compared to its quantum counterpart, its benefit to image quality saturates as it increases [6, 50]. In this work, due to the high-intensity illumination approximation, we are precisely in the aforementioned regime. As shown in Eq. (3), $u$ does not appear and thus does not affect the CNR. Therefore, the impact of the total photon number can be analyzed solely through $K$. When $K = K'$, the advantage of our scheme depends on the fraction of transparent area in the sample. The higher the proportion of transparent regions, the better the CNR. As long as $\alpha \leq 3$, our scheme consistently outperforms traditional methods. In case where transparent and opaque areas are equal ($\alpha = 1$), the CNR of our scheme can theoretically be improved by a factor of $\sqrt{2}$. The above results have an intuitive physical explanation: the very light that our scheme can utilize—light that conventional methods lose through absorption—carries inherent noise. This causes the CNR of our scheme to deteriorate relative to conventional approaches as the sample's opaque area grows.

The above analysis treats the total photon number in the illumination field as a constrained resource (thus keeping $K$ identical in the comparison). Next, we show that relaxing this constraint allows our scheme to achieve significantly enhanced image quality.

## III DISSCUSSION AND ANALYSIS ON IMPROVING IMAGING QUALITY

### A. Enhancing CNR by increasing $K$ under fixed sample absorption

Maintaining the light dose delivered to the sample within a safety threshold is crucial for applications such as in vivo biological imaging. In traditional ghost imaging schemes, the total light absorption by the sample under ideal conditions is given by $I'_{abs} = K'J_b u$. In our scheme, due to multiple interactions between the sample and the light field, the total light absorption over $M$ interactions must be accumulated. Assuming the operators $s^{\dagger(m)}$ and $a^\dagger$ are related by the coefficient $\chi_{abs}^{(m)}$, i.e., $s^{\dagger(m)}(x_b) = \chi_{abs}^{(m)} a^\dagger(x_b)/\sqrt{2}$, the total light absorption throughout the imaging process is:

$$I_{abs} = KJ_b \sum_{m=1}^{M} \left\langle s^{\dagger(m)} s^{(m)} \right\rangle = uKJ_b \sum_{m=1}^{M} \left| \chi_{abs}^{(m)} \right|^2 . \tag{4}$$

The calculation of $\chi_{abs}^{(m)}$ is analogous to that of $\chi_p$ and $\chi_b$. Specifically, for a lossless chain structure with large $M$, we obtain $I_{abs} \approx KJ_b\pi^2 u/4M$ (see Appendix B3), which indicates that the light absorption per measurement in our scheme is far lower than that in the traditional approach, and thus the total absorption is significantly reduced. As $M \to \infty$, the absorption approaches zero. Consequently, even under intense illumination, our scheme ensures that the sample remains undamaged. The underlying principle of this phenomenon is analogous to the quantum Zeno effect: continuous observation (via sample absorption) prevents the light field from transitioning out of its initial path (the lower path), thereby significantly reducing the light dose delivered to the sample. It should be noted that for practical applications, it is not necessary to strictly ensure that the sample absorbs no photons during the measurement process (i.e., with $M$ being infinite). Considering the condition of finite $M$, we refer to this as a quantum Zeno-like effect. An increase in $M$ enhances the sample's tolerance to light-induced damage.

Under the same absorbed light dose as in the traditional scheme, our scheme achieves a photon-number amplification factor of $K/K' = 4M/\pi^2$. This corresponds to an approximate CNR ratio of $\mathrm{CNR}/\mathrm{CNR}' \approx 4\sqrt{M}/(\pi\sqrt{1+\alpha})$. Consequently, our scheme outperforms the traditional method in image quality whenever $M > \alpha$, and the image quality scales proportionally with $\sqrt{M}$.

### B. Enhancing CNR by exploiting controllable optical loss for noise suppression

Prior analysis assumed negligible optical loss in the chain structure. Although losses are generally thought to impair imaging by reducing interference, we next show that controlled optical loss can actually enhance performance.

To simplify the analysis, we first consider the case of large $K$, where Eq. (3) reduces to: $\mathrm{CNR} \approx \sqrt{K/J_p}\,(C_p + C_b)/\sqrt{C_p^2 + \alpha C_b^2}$. By differentiating the CNR with respect to the reflection coefficient $\eta_0$ of the lower optical path and substituting the relation $C_b = \eta_0^{2M}\cos^{2M}\theta$, the extremum condition is found to be either $C_p = \alpha C_b$ or $\eta_0\,\partial C_p/\partial\eta_0 = 2MC_p$。 Due to the typically large value of $M$, the latter condition is generally difficult to satisfy. The former condition $C_p = \alpha C_b$, however, leads to a vanishing photon number difference between the two bucket detectors per measurement, since $\langle I_1\rangle - \langle I_0\rangle = J_p[C_p - \alpha C_b]u$ (see Appendix C). As shown in Appendix C, the background noise difference between the two measurement channels is proportional to this photon-number difference. Thus, $C_p = \alpha C_b$ corresponds to automatic cancellation of background noise. Under this condition and when $\alpha > 1$, the maximum CNR is given by:

$$\mathrm{CNR}_{\mathrm{MAX}} = \sqrt{\frac{K}{J_p}}\sqrt{1+\frac{1}{\alpha}}\,. \tag{5}$$

Evidently, even when $K = K'$, the CNR upper bound in our scheme is always higher than in the traditional scheme, and $\alpha$ ceases to be a limiting factor.

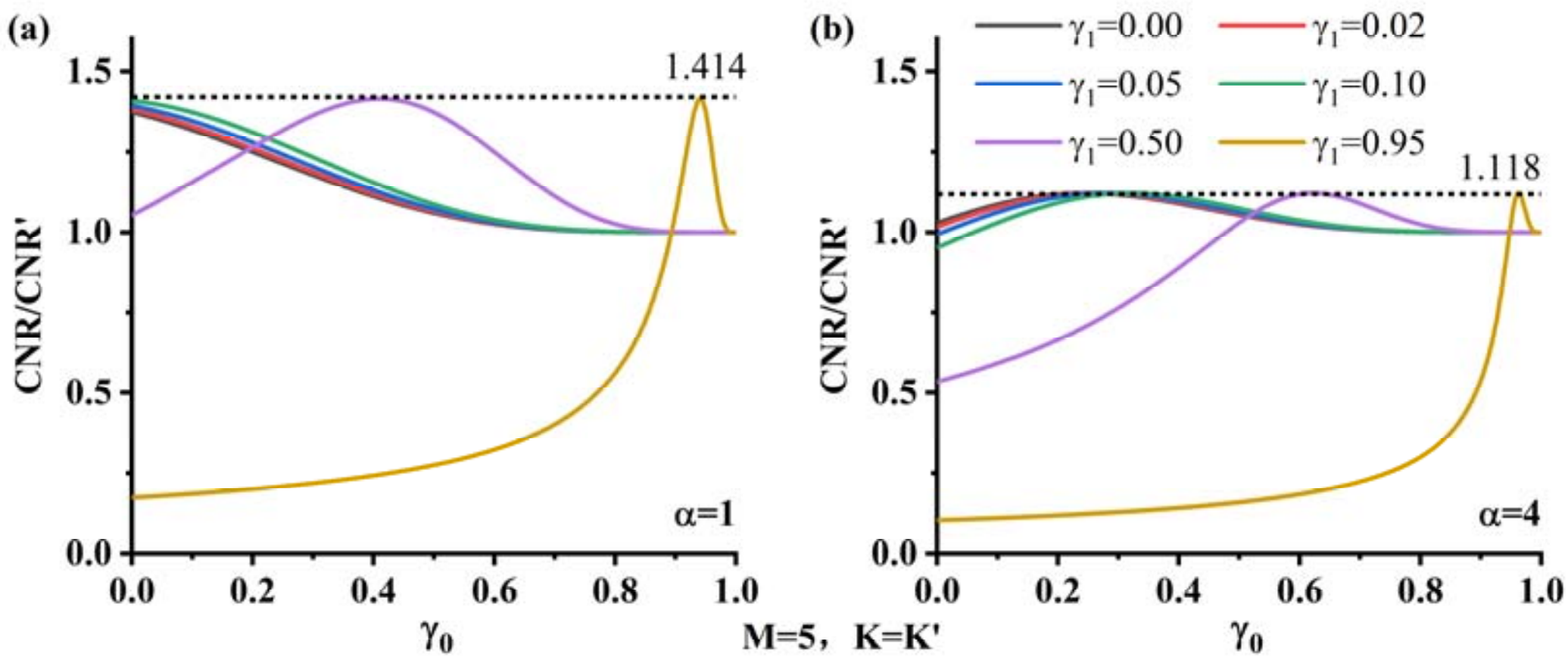

Fig. 2 Ratio of the CNR for the interaction-free scheme to that of the traditional scheme under the condition $K = K'$, plotted as a function of the optical loss probability $\gamma_0$ for different $\gamma_1$. Panels (a) and (b) correspond to $\alpha = 1$ and $\alpha = 4$, respectively.

While the above results establish the theoretical CNR upper limit for our scheme, its practical advantage range under specific optical loss conditions is delineated in Fig. 2. This figure illustrates the ratio of the CNR in our scheme to that of the traditional scheme as a function of the lower-path mirror loss $\gamma_0$ under the condition $K = K'$, with different colored curves representing varying levels of upper-path mirror loss $\gamma_1$. From Fig. 2, not only can an extremum in the CNR be observed, but it is also evident that over a wide range of $\gamma_0$ and $\gamma_1$ values, our scheme maintains a CNR advantage over the traditional scheme (i.e., the ratio is greater than 1). This means that optical losses in the optical chain not only do not degrade imaging performance but actually enhance image quality by suppressing background noise.

To further quantitatively assess the performance gain under a fixed sample absorption, we plot in Fig. 3 the CNR ratio between our scheme and the traditional scheme (assuming an ideal implementation) as a function of $\gamma_0$ and $\gamma_1$. In the simulation, we first numerically calculate the absorption coefficient $\sum_{m=1}^{M} \left|\chi_{abs}^{(m)}\right|^2$, and then determine the required number of measurements for our scheme based on the relation $K = K' / \sum_{m=1}^{M} \left|\chi_{abs}^{(m)}\right|^2$. The results demonstrate that over a broad range of parameters, our scheme substantially outperforms the conventional approach. Only the black regions correspond to no advantage (CNR ratio $\leq 1$). The figure also indicates that the imaging performance improves with increasing $M$. Especially, as shown in Fig. 3(b), when $M = 10$, the CNR of our scheme reaches 10 times that of the conventional scheme. Additionally, as shown in Figs. 3(e) and 3(f), even when $\alpha > 3$, our scheme can still achieve a superior CNR by tuning the optical loss. Although the improvement is less pronounced than when $\alpha$ is small, a performance enhancement of over threefold can still be achieved across a wide range of $\gamma_0$ and $\gamma_1$.

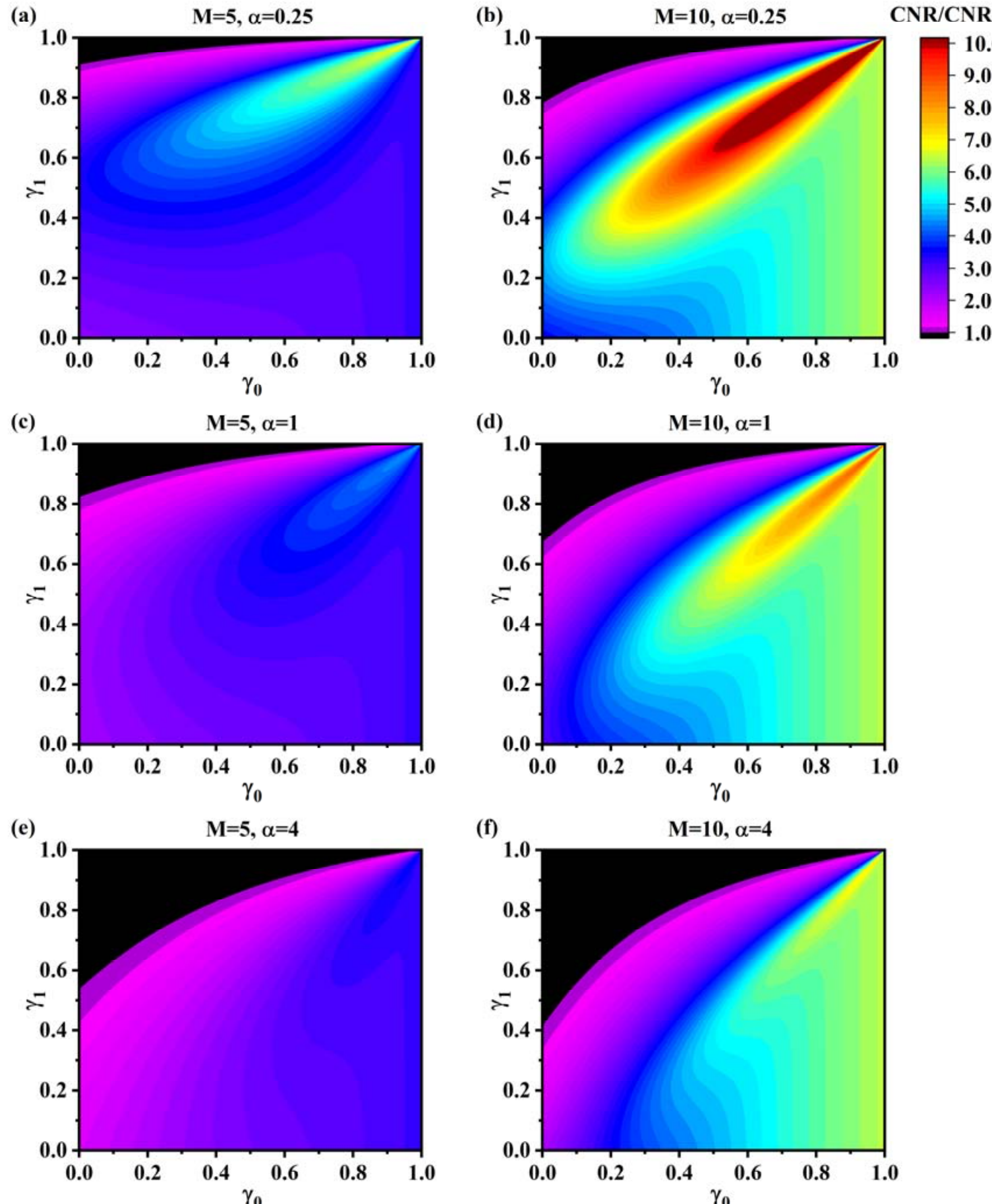

Fig. 3 Ratio of the CNR for the interaction-free scheme to that of the traditional scheme under identical absorption conditions, plotted as a function of optical loss probabilities $\gamma_0$ and $\gamma_1$. The parameters ($M$, $\alpha$) for panels (a-f) are: (a) (5, 0.25), (b) (10, 0.25), (c) (5, 1), (d) (10, 1), (e) (5, 4), and (f) (10, 4).

### C. Validity of the approximation of high-intensity illumination

Although numerical analysis confirms that increasing optical loss can enhance CNR, this enhancement is inherently limited—when optical loss continues to increase, the bucket detector will eventually fail to receive enough photons. This raises a critical issue: once the average number of photons received by the detector decreases to the point where shot noise cannot be ignored, our previous calculations are no longer valid. We therefore proceed to analyze the valid range of optical loss next.

As a reference, we first recall the experimental conditions in [50]: the bucket detector received a light pulse of approximately $1\, nJ$, the sample transparent area was $5\, mm^2$, and the $CCD$ single-pixel area was $10\, \mu m^2$. According to [50], each $CCD$ pixel received about $7800$ photons per pulse, while the bucket detector received approximately $3.9 \times 10^9$ photons. Moreover, [50] indicates that when the number of photons received per pixel exceeds $10$, shot noise can be neglected. Noting that our scheme mainly introduces a chain structure into the signal path compared to [50], we can assume the use of the same light source as in [50] and determine whether shot noise is negligible by comparing the number of photons received by the detectors in the two schemes.

For the $CCD$, since our scheme only adds a 50:50 beam splitter and no additional optical loss is introduced, the shot noise remains negligible. As for the bucket detectors, we note that the average number of photons per measurement in [50] is $\langle I_0' \rangle = J_p\, u$. In our scheme, however, for $i = 0{,}1$, the corresponding average photon number at each bucket detector is $\langle I_i \rangle = \left( J_p |\chi_{pi}|^2 + J_b |\chi_{bi}|^2 \right) u$. Fig. 4 shows the ratio $\langle I_i \rangle / \langle I_0' \rangle$ as a function of optical loss. White curves for ratios of $10^{-2}$, $10^{-4}$, and $10^{-6}$ are also plotted (corresponding to the bucket detectors in our scheme receiving $3.9 \times 10^7$, $3.9 \times 10^5$, and $3.9 \times 10^3$ photons, respectively). By comparing Figs. 3(c), 3(d) and Fig. 4, we observe that even with significant optical attenuation in the chain structure, the detectors still capture enough photons that shot noise becomes negligible, particularly in parameter regimes where our scheme retains a significant CNR advantage, demonstrating the feasibility of improving image quality through optical loss control.

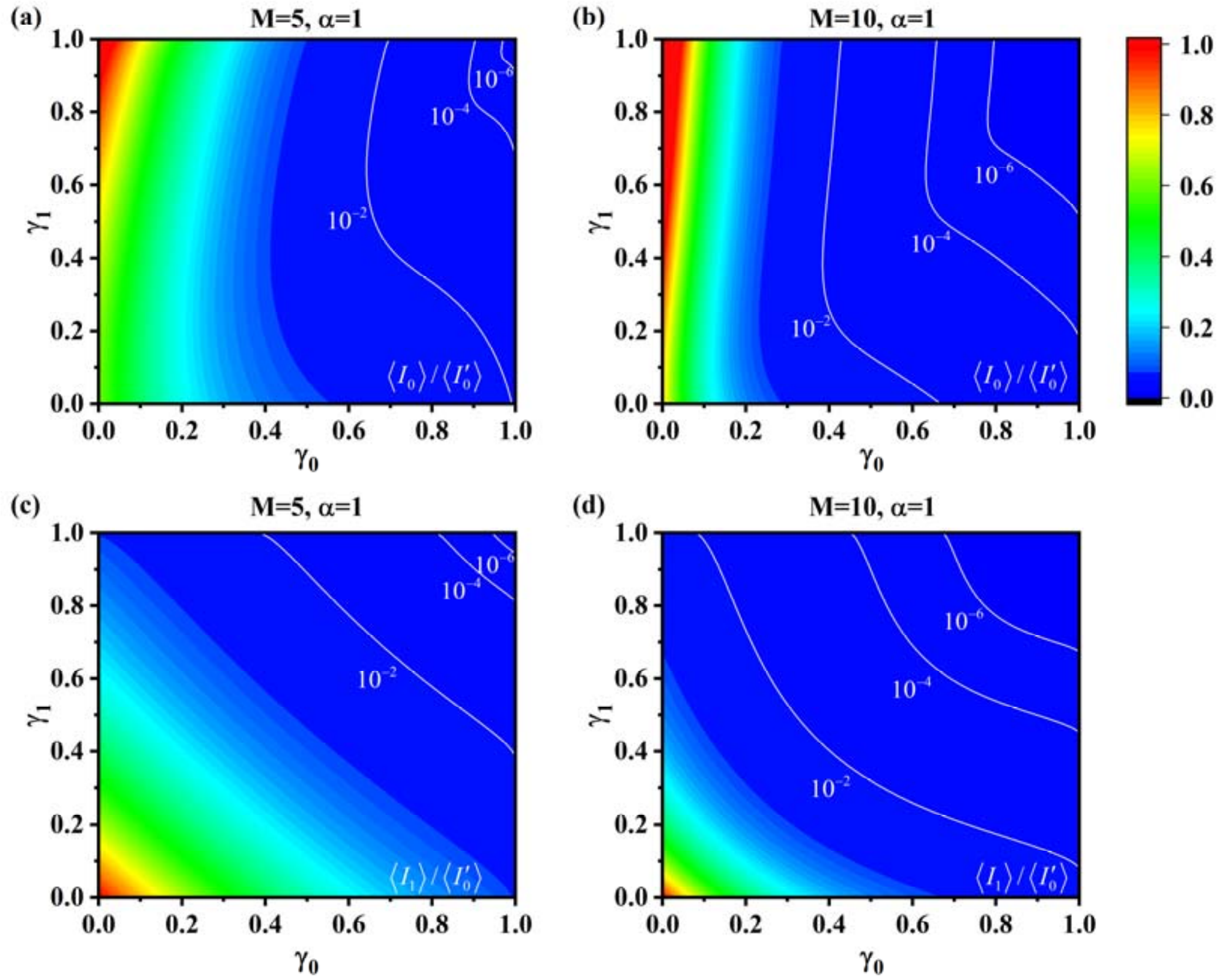

Fig. 4 Comparison of the average number of photons received by the bucket detectors per measurement in the proposed and traditional schemes for different optical loss probabilities $\gamma_0$ and $\gamma_1$, with $\alpha = 1$. Here, $\langle I_0\rangle$, $\langle I_1\rangle$, $\langle I_0'\rangle$ denote the photon numbers collected by $D_0$ (proposed scheme), $D_1$ (proposed scheme), and the bucket detector in the traditional scheme, respectively. The plots show (a) $\langle I_0\rangle/\langle I_0'\rangle$ for $M = 5$, (b) $\langle I_0\rangle/\langle I_0'\rangle$ for $M = 10$, (c) $\langle I_1\rangle/\langle I_0'\rangle$ for $M = 5$, and (d) $\langle I_1\rangle/\langle I_0'\rangle$ for $M = 10$.

## D. VISIBILITY

We now complement our analysis by adopting traditional visibility, which is defined as:

$$V \equiv \frac{\left|\left\langle G\left(x_p\right)\right\rangle - \left\langle G\left(x_b\right)\right\rangle\right|}{\left\langle g_1\left(x_p\right)\right\rangle + \left\langle g_1\left(x_b\right)\right\rangle + \left\langle g_0\left(x_p\right)\right\rangle + \left\langle g_0\left(x_b\right)\right\rangle} = \frac{\left|C_p + C_b\right|}{\left|\chi_{p1}\right|^2 + \left|\chi_{b1}\right|^2 + \left|\chi_{p0}\right|^2 + \left|\chi_{b0}\right|^2}, \tag{6}$$

where the denominator represents the sum of signals from both transparent and opaque sample units in the two sets of $CCD$-bucket detector coincidence measurements. Under ideal conditions, the visibility value is 1. It is worth emphasizing that this normalized visibility depends solely on the parameters of the chain structure itself, such as $M$, $\gamma_0$, and $\gamma_1$, and is independent of other variables including $K$, $u$, and $\alpha$.

Fig. 5 shows the variation of visibility against optical loss probabilities $\gamma_0$ and $\gamma_1$. Fig. 5(a) corresponds to $M = 5$, and Fig. 5(b) corresponds to $M = 10$. The black areas in the figure indicate regions where visibility is below 0.5. Simulation results show that even under high optical loss conditions, the system can still maintain high visibility, provided that $\gamma_0 \approx \gamma_1$.

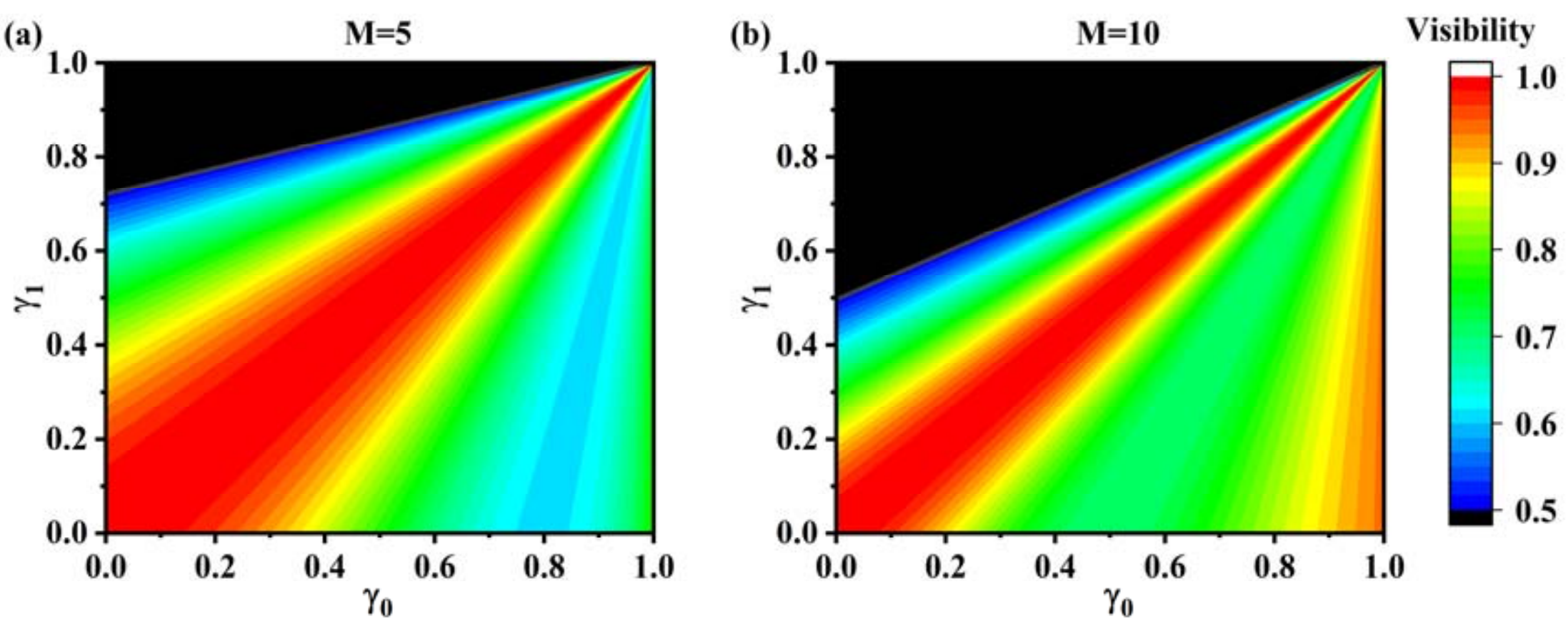

Fig. 5 Visibility versus optical loss probabilities $\gamma_0$ and $\gamma_1$ with (a) $M = 5$ and (b) $M = 10$.

## IV. CONCLUSION

We propose an interaction-free ghost imaging scheme based on a thermal light source, aiming to provide an alternative approach to high-quality, low-damage, and cost-effective imaging of light-sensitive objects, featuring three key advantages. First, the chain interferometer structure in the signal path not only recycles photons that would otherwise be absorbed in conventional schemes to improve image quality, but also induces a quantum Zeno-like effect that significantly reduces optical damage—thereby making the sample nearly immune to light-induced damage as $M$ increases. This reduction enables a safe and substantial increase in measurement repetitions, thereby significantly enhancing image quality. Second, we demonstrate that optical loss in the chain structure can serve as an effective means to suppress background noise, ensuring that the upper limit of our scheme's CNR is higher than that of traditional thermal ghost imaging. Third, our scheme eliminates the need for quantum light sources and single-photon detectors, enabling faster, more robust, and lower-cost imaging.

## ACKNOWLEDGMENTS

This work is sponsored by Natural Science Foundation of Shanghai (Grant Nos. 25ZR1401140 and 25ZR1401114). This work is also supported by National Natural Science

Foundation of China (Grant No. 62161038). Z.-H.L thanks M. Suhail Zubairy for discussions on interaction-free measurement with thermal light.

## APPENDIX A: VACUUM-INPUT-NEGLECTING APPROXIMATION

Consider an arbitrary linear optical system with $N$ inputs and $N$ outputs, where the input optical fields are described by operators $a_n^\dagger$ and the output field operators by $d_n^\dagger$ with $n \in [0, N)$. We demonstrate that for the coincidence measurement between any two outputs, the influence of all vacuum inputs can be neglected, provided that the thermal light input has an average photon number much greater than 1.

For convenience, we assume that only the input corresponding to operator $a_0^\dagger$ is non-zero, while all other inputs are in vacuum states. In other words, supposing that the probability of the initial state $|\varphi_i\rangle$ appearing is $P_i$, we have:

$$a_{n\neq 0}(x)\left|\varphi_i\right\rangle = 0\,, \tag{A1}$$

$$\sum_{i=0}^{\infty} P_i \left\langle \varphi_i \right| a_0^\dagger(x) a_0(x) \left|\varphi_i\right\rangle = I\,, \tag{A2}$$

where $x$ represents different positions on the input/output plane, and $I$ is the average number of photons in the incident field.

Because the system is linear, we can relate the initial state $|\varphi_i\rangle$ and the final state $|\phi_i\rangle$ of the optical field through a time evolution operator $U$, i.e.,

$$U\left|\varphi_i\right\rangle = \left|\phi_i\right\rangle. \tag{A3}$$

The evolution of the corresponding operators is assumed to take the following form,

$$U^\dagger d_n^\dagger(x) U = \sum_{l=0}^{N-1} \chi_{n,l}(x) a_l^\dagger(x)\,, \tag{A4}$$

where $\chi_{n.l}$ is the transformation coefficient satisfying the normalization condition $\sum_{l=0}^{N-1} \left|\chi_{n,l}(x)\right|^2 = 1$.

We now consider the coincidence measurement between two outputs corresponding to the operators $d_n^\dagger$ and $d_{n'}^\dagger$ as follows,

$$\left\langle \phi_i \right| d_n^\dagger(x) d_n(x) d_{n'}^\dagger(x') d_{n'}(x') \left|\phi_i\right\rangle. \tag{A5}$$

Using Eqs. A3 and A4, Eq. A5 becomes,

$$\sum_{l,l',l'',l'''=0}^{N-1} \chi_{n,l}(x) \chi_{n,l'}^*(x) \chi_{n',l''}(x') \chi_{n',l'''}^*(x') \left\langle \varphi_i \right| a_l^\dagger(x) a_{l'}(x) a_{l''}^\dagger(x') a_{l'''}(x') \left|\varphi_i\right\rangle. \tag{A6}$$

By converting Eq. A6 to normal-ordered form, we obtain

$$\begin{aligned}&\sum_{l,l',l'''=0}^{N-1} \chi_{n,l}(x)\chi^*_{n,l'}(x)\chi_{n',l'}(x')\chi^*_{n',l'''}(x')\left\langle \varphi_i \middle| a_l^\dagger(x)a_{l'''}(x') \middle| \varphi_i \right\rangle \delta(x-x') \\ &+\sum_{l,l',l'',l'''=0}^{N-1} \chi_{n,l}(x)\chi^*_{n,l'}(x)\chi_{n',l''}(x')\chi^*_{n',l'''}(x')\left\langle \varphi_i \middle| a_l^\dagger(x)a_{l''}^\dagger(x')a_{l'}(x)a_{l'''}(x') \middle| \varphi_i \right\rangle.\end{aligned} \tag{A7}$$

Using Eq. A1, we can eliminate all non-zero terms, which simplifies Eq.A7 to

$$\begin{aligned}&\sum_{l'=0}^{N-1} \chi_{n,0}(x)\chi^*_{n,l'}(x)\chi_{n',l'}(x')\chi^*_{n',0}(x')\left\langle \varphi_i \middle| a_0^\dagger(x)a_0(x') \middle| \varphi_i \right\rangle \delta(x-x') \\ &+\chi_{n,0}(x)\chi^*_{n,0}(x)\chi_{n',0}(x')\chi^*_{n',0}(x')\left\langle \varphi_i \middle| a_0^\dagger(x)a_0^\dagger(x')a_0(x)a_0(x') \middle| \varphi_i \right\rangle.\end{aligned} \tag{A8}$$

Next, we account for the influence of all possible initial states of the system. The expectation value of Eq. A5 becomes,

$$\begin{aligned}&\sum_{i=0}^{\infty} P_i \left\langle \phi_i \middle| d_n^\dagger(x)d_n(x)d_{n'}^\dagger(x')d_{n'}(x') \middle| \phi_i \right\rangle \\ &=\sum_{l'=0}^{N-1} \chi_{n,0}(x)\chi^*_{n,l'}(x)\chi_{n',l'}(x')\chi^*_{n',0}(x')\left\langle a_0^\dagger(x)a_0(x') \right\rangle \delta(x-x') \\ &+\chi_{n,0}(x)\chi^*_{n,0}(x)\chi_{n',0}(x')\chi^*_{n',0}(x')\left\langle a_0^\dagger(x)a_0^\dagger(x')a_0(x)a_0(x') \right\rangle.\end{aligned} \tag{A9}$$

Assuming the input is thermal, we apply the Gaussian moment theorem, which yields:

$$\begin{aligned}&\chi_{n,0}(x)\chi^*_{n',0}(x')\left\langle a_0^\dagger(x)a_0(x') \right\rangle \delta(x-x')\sum_{l'=0}^{N-1}\chi^*_{n,l'}(x)\chi_{n',l'}(x') \\ &+\chi_{n,0}(x)\chi^*_{n,0}(x)\chi_{n',0}(x')\chi^*_{n',0}(x') \\ &\times\left[\left\langle a_0^\dagger(x)a_0(x) \right\rangle\left\langle a_0^\dagger(x')a_0(x') \right\rangle+\left\langle a_0^\dagger(x)a_0(x') \right\rangle\left\langle a_0^\dagger(x')a_0(x) \right\rangle\right].\end{aligned} \tag{A10}$$

We proceed to analyze Eq. A10 in two distinct cases. Utilizing Eq. A2, we obtain,

Case1： if $x = x'$， Eq. A10 becomes,

$$I\chi_{n,0}(x)\chi^*_{n',0}(x)\sum_{l'=0}^{N-1}\chi^*_{n,l'}(x)\chi_{n',l'}(x)+2I^2\chi_{n,0}(x)\chi^*_{n,0}(x)\chi_{n',0}(x)\chi^*_{n',0}(x). \tag{A11}$$

Case2： if $x \neq x'$， Eq. A10 becomes,

$$I^2\chi_{n,0}(x)\chi^*_{n,0}(x)\chi_{n',0}(x')\chi^*_{n',0}(x'). \tag{A12}$$

Eq. S1.11 demonstrates that the contribution from vacuum inputs (the first term $\sum_{l'=0}^{N-1}\chi^*_{n,l'}(x)\chi_{n',l'}(x)$) is one order of magnitude smaller in $I$ than the term corresponding to the thermal light input (the second term in Eq. A11). In thermal light imaging, the average photon number of the illumination field per measurement is generally much greater than 1, rendering the influence of vacuum inputs negligible. While the above discussion focuses on second-order correlation functions, similar conclusions hold for higher-order correlation functions. Consequently, in the calculations presented in this work, for any output creation operator $d^{\dagger}$, if it cannot originate from any non-vacuum input, we explicitly set $d^{\dagger}=0$.

## APPENDIX B: CALCULATION OF THE INPUT-OUTPUT OPERATOR RELATIONS FOR A LOSSLESS CHAIN STRUCTURE

Here, we progressively derive the operator relations between the system's outputs and inputs by tracing the reverse evolution of the optical field from the detector back to the source, accounting for each optical component encountered in the process.

### B1 Case of a Transparent Sample Unit

During the reverse evolution process, the optical field first encounters $BS_M^{(M)}$. Using the input-output relation for a beam splitter given in the main text, we have:

$$d_0^{\dagger(M)}=d_0^{\dagger(M-1)}\cos\theta-d_1^{\dagger(M-1)}\sin\theta\,, \qquad \text{B1}$$

$$d_1^{\dagger(M)}=d_0^{\dagger(M-1)}\sin\theta+d_1^{\dagger(M-1)}\cos\theta\,. \qquad \text{B2}$$

Since the output and input planes have a one-to-one positional correspondence, we omit the spatial coordinate $x$ here (this convention will be followed throughout the subsequent discussion).

After passing through $BS_M^{(M)}$, the optical field will subsequently encounter $BS_M^{(M-1)}$. Using the input-output operator relation for $BS_M^{(M-1)}$, we can further derive the relation between $d^{\dagger(M)}$ and $d^{\dagger(M-2)}$:

$$d_0^{\dagger(M)}=d_0^{\dagger(M-2)}\cos 2\theta-d_1^{\dagger(M-2)}\sin 2\theta\,, \qquad \text{B3}$$

$$d_1^{\dagger(M)}=d_0^{\dagger(M-2)}\sin 2\theta+d_1^{\dagger(M-2)}\cos 2\theta\,. \qquad \text{B4}$$

This process will repeat iteratively. We can obtain the relation between the input operator at $BS_M^{(1)}$ and the output operator at $BS_M^{(M)}$ as follows:

$$d_0^{\dagger(M)}=d_0^{\dagger(0)}\cos M\theta-d_1^{\dagger(0)}\sin M\theta=-d_1^{\dagger(0)}\,, \qquad \text{B5}$$

$$d_1^{\dagger(M)}=d_0^{\dagger(0)}\sin M\theta+d_1^{\dagger(0)}\cos M\theta=d_0^{\dagger(0)}\,. \qquad \text{B6}$$

In the second equality of Eqs. B5 and B6, we have applied the condition $\theta=\pi/2M$.

According to our explanation in Appendix A, since the optical field corresponding to $d_0^{\dagger(M)}$ cannot originate from the thermal light source, we set:

$$d_0^{\dagger(M)}=0\,. \qquad \text{B7}$$

Note that we have the following definition in the main text:

$$d_{i=0,1}^{\dagger(M)}(x_p)=\frac{1}{\sqrt{2}}\chi_{pi}a^{\dagger}(x_p)\,, \quad \text{B8}$$

where the coefficient $1/\sqrt{2}$ originates from the 50:50 beam splitter $BS_H^{(1)}$. Based on this definition and combining Eqs. A6 and A7, we obtain the relation given in the main text:

$$\chi_{p0}=0\,, \quad \text{B9}$$

$$\chi_{p1}=1\,. \quad \text{B10}$$

**B2 Case of an Opaque Sample Unit**

Photon absorption by the sample unit necessarily alters the evolution of the optical field. Taking the optical field corresponding to operator $d_1^{\dagger(M)}$ as an example, since the sample absorbs all photons in the path, this implies that such field cannot originate from the light source. Therefore, we can set:

$$d_1^{\dagger(M)}=0\,. \quad \text{B11}$$

For the optical field corresponding to operator $d_0^{\dagger(M)}$, considering the effect of $BS_M^{(M)}$, we have:

$$d_0^{\dagger(M)}=d_0^{\dagger(M-1)}\cos\theta-d_1^{\dagger(M-1)}\sin\theta\,. \quad \text{B12}$$

Next, we proceed to analyze the optical fields corresponding to operators $d_1^{\dagger(M-1)}$ and $d_0^{\dagger(M-1)}$. Similarly, due to the absorption effect of the sample, for $d_1^{\dagger(M-1)}$ we have:

$$d_1^{\dagger(M-1)}=0\,. \quad \text{B13}$$

By substituting Eq. B13 and the relation between the input and output operators of $BS_M^{(M-1)}$ into Eq. B12, we obtain:

$$d_0^{\dagger(M)}=\left[d_0^{\dagger(M-2)}\cos\theta-d_1^{\dagger(M-2)}\sin\theta\right]\cos\theta\,. \quad \text{B14}$$

The above process repeats iteratively. Finally, we obtain the relation between the input operator of $BS_M^{(1)}$ and the output operator of $BS_M^{(M)}$:

$$d_0^{\dagger(M)}=\left[d_0^{\dagger(0)}\cos\theta-d_1^{\dagger(0)}\sin\theta\right]\cos^{M-1}\theta\,. \quad \text{B15}$$

Noting that $d_1^{\dagger(0)}$ corresponds to a vacuum input field, i.e., $d_1^{\dagger(0)}=0$, Eq. B15 can be rewritten as,

$$d_0^{\dagger(M)}=d_0^{\dagger(0)}\cos^{M}\theta\,. \quad \text{B16}$$

Combining Eq. B11 with the following definition provided in the main text,

$$d_{i=0,1}^{\dagger(M)}(x_b)=\frac{1}{\sqrt{2}}\chi_{bi}a^{\dagger}(x_b)\,, \quad \text{B17}$$

we have,

$$\chi_{b1}=0\,, \quad \text{B18}$$

$$\chi_{b0}=\cos^{M}\theta\,. \quad \text{B19}$$

Since $\theta=\pi/2M$, it can be seen that as $M\to\infty$, we have $\chi_{b0}\to 1$. Additionally, analogous to the calculations above, we can conclude that for an opaque sample with optical loss in the chain structure, loss in the upper path does not affect the result, whereas the effect of loss in the lower path leads to:

$$\chi_{b0}=\eta_0^{M}\cos^{M}\theta\,, \quad \text{B20}$$

$$\chi_{b1}=0\,. \quad \text{B21}$$

**B3 Total Light Absorption of the Sample**

We first consider the operator $s^{\dagger(M)}$. The relevant calculation essentially involves tracing the reverse evolution of the optical field starting from the sample located between $BS_M^{(M)}$ and $D_1$ in Fig. 1(a) of the main text. Since the sample unit is in an opaque state, the calculation is analogous to that in Appendix B2, and thus we have:

$$s^{\dagger(M)}=d_0^{\dagger(0)}\cos^{M-1}\theta\sin\theta\,. \quad \text{B22}$$

For $s^{\dagger(M-1)}$, the calculation method is similar, except that the starting point now becomes the sample between $BS_M^{(M-1)}$ and $BS_M^{(M)}$. The entire calculation process is analogous to that for $d_0^{\dagger(M-1)}$, and therefore we have:

$$s^{\dagger(M-1)}=d_0^{\dagger(0)}\cos^{M-2}\theta\sin\theta\,. \quad \text{B23}$$

The description of the sample's light absorption in a single measurement must include all $s^{\dagger(m)}$ $(m=1,2,\dots,M)$, where

$$s^{\dagger(m)}=d_0^{\dagger(0)}\cos^{m-1}\theta\sin\theta\,. \quad \text{B24}$$

Correspondingly, the total light absorption of the sample is:

$$\begin{aligned}I_{abs}&=KJ_b\sum_{m=1}^{M}\left\langle s^{\dagger(m)}s^{(m)}\right\rangle\\&=KJ_b\sum_{m=1}^{M}\cos^{2m-2}\theta\sin^2\theta\left\langle d_0^{\dagger(0)}d_0^{(0)}\right\rangle.\\&=uKJ_b\left(1-\cos^{2M}\theta\right).\end{aligned} \quad \text{B25}$$

When $M\to\infty$, we can obtain that $I_{abs}\approx uKJ_b\pi^2/4M$.

## APPENDIX C: CALCULATION OF CNR AND VISIBILITY

To facilitate the derivation, we define:

$$I_r(x) = b^\dagger(x)b(x)\,. \tag{C1}$$

Note that both beam splitters in the reference path are 50:50 beam splitters, therefore we have

$$c^\dagger_{i=0,1}(x) = \frac{1}{\sqrt{2}}b^\dagger(x)\,, \tag{C2}$$

$$b^\dagger(x) = \frac{1}{\sqrt{2}}a^\dagger(x)\,. \tag{C3}$$

Combining Eq. C2 with Eq. C1, we have

$$I_r(x) = 2I_{ci}(x)\,. \tag{C4}$$

Thus, the expression for $g_i(x)$ in the main text can be updated to

$$g_i(x) = \frac{1}{2}\left\{E\left[I_i I_r(x)\right] - E\left[I_i\right]E\left[I_r(x)\right]\right\}. \tag{C5}$$

Based on the definition of function $E[\ldots]$, we calculate the average over $K$ independent measurements. The derivation yields

$$\left\langle E\left[I_i I_r(x)\right]\right\rangle = \frac{1}{K}\sum_{k=1}^{K} \left\langle I_i^{(k)} I_r^{(k)}(x)\right\rangle = \left\langle I_i I_r(x)\right\rangle, \tag{C6}$$

$$\begin{aligned}\left\langle E\left[I_i\right]E\left[I_r(x)\right]\right\rangle &= \left\langle \left[\frac{1}{K}\sum_{k=1}^{K} I_i^{(k)}\right]\left[\frac{1}{K}\sum_{k'=1}^{K} I_r^{(k')}(x)\right]\right\rangle \\ &= \frac{1}{K^2}\sum_{k=1}^{K} \left\langle I_i^{(k)} I_r^{(k)}(x)\right\rangle + \frac{1}{K^2}\sum_{k=1}^{K}\sum_{\substack{k'=1\\k'\neq k}}^{K} \left\langle I_i^{(k)}\right\rangle\left\langle I_r^{(k')}(x)\right\rangle \\ &= \frac{1}{K}\left\langle I_i I_r(x)\right\rangle + \frac{K-1}{K}\left\langle I_i\right\rangle\left\langle I_r(x)\right\rangle.\end{aligned} \tag{C7}$$

Here, the superscript in parentheses $k$ represents the result of the $k$-th independent measurement. We have assumed that the expectation value is identical for each independent measurement. Additionally, it is worth noting that Eq. C7 shows that when $K$ is large, the background noise in the coincidence measurement between the bucket detector $D_i$ and the $CCD_i$ can be approximated as:

$$\left\langle I_i\right\rangle\left\langle I_r(x)\right\rangle = \frac{1}{2}\left\langle I_i\right\rangle\left\langle I_a(x)\right\rangle = \left\langle I_i\right\rangle u\,. \tag{C8}$$

The background noise difference between the two detection channels is thus given by:

$$\left\langle I_1\right\rangle\left\langle I_r(x)\right\rangle-\left\langle I_0\right\rangle\left\langle I_r(x)\right\rangle=\left(\left\langle I_1\right\rangle-\left\langle I_0\right\rangle\right)u\ . \tag{C9}$$

In other words, the background noise difference is determined by the average numbers of photons received by the two bucket detectors.

Substituting Eqs. C6 and C7 into Eq. C5, we obtain

$$\left\langle g_i\left(x\right)\right\rangle=\frac{K-1}{2K}\left[\left\langle I_i I_r\left(x\right)\right\rangle-\left\langle I_i\right\rangle\left\langle I_r(x)\right\rangle\right]. \tag{C10}$$

Following a similar calculation to Eq. C10, we can also obtain

$$\begin{aligned}
&\left\langle g_i(x)g_{i'}(x)\right\rangle\\
&=\frac{\left(K^2-2K+2\right)(K-1)}{4K^3}\left\langle I_i I_r(x)\right\rangle\left\langle I_{i'} I_r(x)\right\rangle\\
&+\frac{(K-1)(K-2)(K-3)}{4K^3}\left\langle I_i\right\rangle\left\langle I_{i'}\right\rangle\left\langle I_r(x)\right\rangle^2\\
&-\frac{(K-1)(K-2)^2}{4K^3}\left(\left\langle I_i I_r\left(x\right)\right\rangle\left\langle I_{i'}\right\rangle+\left\langle I_{i'} I_r\left(x\right)\right\rangle\left\langle I_i\right\rangle\right)\left\langle I_r(x)\right\rangle\\
&+\frac{\left(K-1\right)^2}{4K^3}\left(\left\langle I_i I_{i'} I_r^2(x)\right\rangle-\left\langle I_i I_r^2(x)\right\rangle\left\langle I_{i'}\right\rangle-\left\langle I_{i'} I_r^2(x)\right\rangle\left\langle I_i\right\rangle-2\left\langle I_i I_{i'} I_r(x)\right\rangle\left\langle I_r(x)\right\rangle\right)\\
&+\frac{(K-1)(K-2)}{4K^3}\left(\left\langle I_i I_{i'}\right\rangle\left\langle I_r(x)\right\rangle^2+\left\langle I_i\right\rangle\left\langle I_{i'}\right\rangle\left\langle I_r^2(x)\right\rangle\right)\\
&+\frac{K-1}{4K^3}\left\langle I_i I_{i'}\right\rangle\left\langle I_r^2(x)\right\rangle,
\end{aligned} \tag{C11}$$

where $i, i' = 0,1$.

Based on the definition of $\langle G(x)\rangle$ provided in the main text, we can further obtain

$$\begin{aligned}
\left\langle G\left(x\right)\right\rangle&=\left\langle g_1(x)\right\rangle-\left\langle g_0(x)\right\rangle\\
&=\frac{K-1}{2K}\left\langle\left(I_1-I_0\right)I_r(x)\right\rangle-\frac{K(K-1)}{2K^2}\left\langle I_1-I_0\right\rangle\left\langle I_r(x)\right\rangle,
\end{aligned} \tag{C12}$$

$$\begin{aligned}
\left\langle G^2\left(x\right)\right\rangle&=\left\langle g_1^2(x)\right\rangle+\left\langle g_0^2(x)\right\rangle-\left\langle g_1(x)g_0(x)\right\rangle-\left\langle g_0(x)g_1(x)\right\rangle\\
&=\frac{K^2-2K+1}{4K^3}\left\langle\left(I_1-I_0\right)^2 I_r^2(x)\right\rangle+\frac{K^3-3K^2+4K-2}{4K^3}\left\langle\left(I_1-I_0\right)I_r(x)\right\rangle^2\\
&-\frac{K^2-2K+1}{2K^3}\left[\left\langle\left(I_1-I_0\right)^2 I_r(x)\right\rangle\left\langle I_r(x)\right\rangle+\left\langle\left(I_1-I_0\right)I_r^2(x)\right\rangle\left\langle I_1-I_0\right\rangle\right]\\
&-\frac{K^3-5K^2+8K-4}{2K^3}\left\langle\left(I_1-I_0\right)I_r\left(x\right)\right\rangle\left\langle I_r(x)\right\rangle\left\langle I_1-I_0\right\rangle\\
&+\frac{K^3-6K^2+11K-6}{4K^3}\left\langle I_1-I_0\right\rangle^2\left\langle I_r(x)\right\rangle^2+\frac{K-1}{2K^3}\left\langle\left(I_1-I_0\right)^2\right\rangle\left\langle I_r^2(x)\right\rangle\\
&+\frac{K^2-3K+2}{4K^3}\left[\left\langle I_1-I_0\right\rangle^2\left\langle I_r^2(x)\right\rangle+\left\langle\left(I_1-I_0\right)^2\right\rangle\left\langle I_r(x)\right\rangle^2\right].
\end{aligned} \tag{C13}$$

In the subsequent calculations, we need to distinguish between the cases where the sample unit is transparent (corresponding to $x_p$) and opaque (corresponding to $x_b$). Taking the first term in Eq. B12, $\langle (I_1 - I_0) I_r(x) \rangle \equiv \langle I_1 I_r(x) \rangle - \langle I_0 I_r(x) \rangle$, as an example and using the results obtained from Appendix B, we can derive:

$$\left\langle I_i I_r\left(x_p\right)\right\rangle = \frac{1}{2^2} \sum_{\substack{x' \in \mathbf{J_p} \\ x' \neq x_p}} \left|\chi_{pi}\right|^2 \left\langle I_a(x')\right\rangle \left\langle I_a(x_p)\right\rangle + \frac{1}{2^2}\left|\chi_{pi}\right|^2 \left\langle I_a(x_p) I_a(x_p)\right\rangle$$

$$+\frac{1}{2^2} \sum_{x' \in \mathbf{J_b}} \left|\chi_{bi}\right|^2 \left\langle I_a(x')\right\rangle \left\langle I_a(x_p)\right\rangle \quad \text{C14}$$

$$= (J_p + 1)\left|\chi_{pi}\right|^2 u^2 + J_b \left|\chi_{bi}\right|^2 u^2,$$

$$\left\langle I_i I_r\left(x_b\right)\right\rangle = J_p \left|\chi_{pi}\right|^2 u^2 + u^2 (J_b + 1)\left|\chi_{bi}\right|^2 . \quad \text{C15}$$

From Eqs. C14 and C15, we can obtain

$$\left\langle \left(I_1 - I_0\right) I_r\left(x_b\right)\right\rangle = \left[C_p J_p - \left(J_b + 1\right) C_b\right] u^2 , \quad \text{C16}$$

$$\left\langle \left(I_1 - I_0\right) I_r\left(x_p\right)\right\rangle = \left[\left(J_p + 1\right) C_p - C_b J_b\right] u^2 . \quad \text{C17}$$

Similarly, the following results can be derived:

$$\left\langle \left(I_1 - I_0\right)^2 I_r^2\left(x_b\right)\right\rangle = 2 J_p \left(1 + J_p\right) u^4 C_p^2 -4 J_p \left(2 + J_b\right) u^4 C_p C_b + 2\left(J_b^2 + 5 J_b + 6\right) u^4 C_b^2 , \quad \text{C18}$$

$$\left\langle \left(I_1 - I_0\right)^2 I_r^2\left(x_p\right)\right\rangle = 2 J_b \left(J_b + 1\right) u^4 C_b^2 -4 J_b \left(2 + J_p\right) u^4 C_p C_b + 2\left[J_p^2 + 5 J_p + 6\right] u^4 C_p^2 , \quad \text{C19}$$

$$\left\langle I_r\left(x_p\right)\right\rangle = \left\langle I_r\left(x_b\right)\right\rangle = u , \quad \text{C20}$$

$$\left\langle I_r^2(x_p)\right\rangle = \left\langle I_r^2(x_p)\right\rangle = 2u^2 , \quad \text{C21}$$

$$\left\langle \left(I_1 - I_0\right) I_r^2\left(x_b\right)\right\rangle = 2u^3 J_p C_p - 2u^3 \left(J_b + 2\right) C_b , \quad \text{C22}$$

$$\left\langle \left(I_1 - I_0\right) I_r^2\left(x_p\right)\right\rangle = 2u^3 \left(2 + J_p\right) C_p - 2u^3 J_b C_b , \quad \text{C23}$$

$$\left\langle \left(I_1 - I_0\right)^2 \right\rangle = C_p^2 u^2 J_p \left(J_p + 1\right) + C_b^2 u^2 J_b \left(J_b + 1\right) - 2 C_p C_b u^2 J_b J_p , \quad \text{C24}$$

$$\left\langle \left(I_1 - I_0\right)^2 I_r\left(x_p\right)\right\rangle = u^3 C_p^2 \left(J_p^2 + 3 J_p + 2\right) -2u^3 C_p C_b J_b \left(J_p + 1\right) + u^3 C_b^2 J_b \left(J_b + 1\right), \quad \text{C25}$$

$$\left\langle \left(I_1 - I_0\right)^2 I_r\left(x_b\right)\right\rangle = C_p^2 u^3 J_p \left(J_p + 1\right) -2u^3 C_p C_b J_p \left(J_b + 1\right) + u^3 C_b^2 \left(J_b^2 + 3 J_b + 2\right), \quad \text{C26}$$

and

$$\left\langle I_{i=0,1}\right\rangle=\left(J_p\left|\chi_{pi}\right|^2+J_b\left|\chi_{bi}\right|^2\right)u\,. \qquad \text{C27}$$

Eq. C27 represents the average photon numbers received by the two bucket detectors per measurement, and their difference is:

$$\left\langle I_1\right\rangle-\left\langle I_0\right\rangle=\left(J_pC_p-J_bC_b\right)u\,. \qquad \text{C28}$$

According to our previous analysis, it can be used to describe the background noise difference between the two detection channels.

Substituting Eqs. C16-C28 into Eq. C.12 yields Eq. (2) in the main text, i.e.,

$$\left\langle G\left(x_p\right)\right\rangle=\frac{K-1}{2K}C_pu^2\,, \qquad \text{C29}$$

$$\left\langle G\left(x_b\right)\right\rangle=-\frac{K-1}{2K}C_bu^2\,. \qquad \text{C30}$$

Substituting Eqs. C16-C28 into Eq. C13 and utilizing $\delta^2G\equiv\langle G^2\rangle-\langle G\rangle^2$, we obtain

$$\delta^2G\left(x_p\right)=\frac{u^4}{4K^3}\left[K\left(K-1\right)\left(J_bC_b^2+J_pC_p^2\right)+\left(7K^2-13K+6\right)C_p^2\right], \qquad \text{C31}$$

$$\delta^2G\left(x_b\right)=\frac{u^4}{4K^3}\left[K\left(K-1\right)\left(J_pC_p^2+J_bC_b^2\right)+\left(7K^2-13K+6\right)C_b^2\right]. \qquad \text{C32}$$

Substituting Eqs. C14, C15, C20 and C27 into Eq. C10 yields the denominator term of the visibility.

$$\begin{aligned}&\left\langle g_1\left(x_p\right)\right\rangle+\left\langle g_1\left(x_b\right)\right\rangle+\left\langle g_0\left(x_p\right)\right\rangle+\left\langle g_0\left(x_b\right)\right\rangle\\&\qquad=\frac{K-1}{2K}\left(\left|\chi_{p1}\right|^2+\left|\chi_{b1}\right|^2+\left|\chi_{p0}\right|^2+\left|\chi_{b0}\right|^2\right)u^2.\end{aligned} \qquad \text{C33}$$

Using Eqs. C29-C33, we finally arrive at the CNR (Eq. (3) in the main text), which is

$$\begin{aligned}CNR&\equiv\frac{\left\langle G(x_p)\right\rangle-\left\langle G(x_b)\right\rangle}{\sqrt{\frac{1}{2}\left[\delta^2G(x_p)+\delta^2G(x_b)\right]}}\\&=\frac{\sqrt{K-1}\left(C_p+C_b\right)}{\sqrt{\left(J_p+\frac{7}{2}-\frac{3}{K}\right)C_p^2+\left(J_b+\frac{7}{2}-\frac{3}{K}\right)C_b^2}}.\end{aligned} \qquad \text{C34}$$

The visibility (Eq. (6) in the main text) is:

$$V \equiv \frac{\left|\left\langle G\left(x_p\right)\right\rangle - \left\langle G\left(x_b\right)\right\rangle\right|}{\left\langle g_1\left(x_p\right)\right\rangle + \left\langle g_1\left(x_b\right)\right\rangle + \left\langle g_0\left(x_p\right)\right\rangle + \left\langle g_0\left(x_b\right)\right\rangle} = \frac{\left|C_p + C_b\right|}{\left|\chi_{p1}\right|^2 + \left|\chi_{b1}\right|^2 + \left|\chi_{p0}\right|^2 + \left|\chi_{b0}\right|^2}. \quad \text{C35}$$